\documentclass[journal=jacsat,manuscript=article]{achemso}

\usepackage{subfiles}
\usepackage{chemformula}
\usepackage[T1]{fontenc}
\usepackage{makecell}
\usepackage[version=4]{mhchem}
\usepackage{xcolor}
\usepackage{soul} 
\usepackage{amsfonts} 
\usepackage{tabularx}
\usepackage{booktabs,makecell}
\usepackage{enumitem}

\author{Yue Yin}
\altaffiliation{These authors contributed equally to this work.}
\author{Jiangshan He}
\altaffiliation{These authors contributed equally to this work.}
\author{Runze Li}
\author{Yunze Qiu}
\author{Dingsheng Wang}
\author{Jun Li}
\author{Hai Xiao}
\email{haixiao@tsinghua.edu.cn}
\affiliation{Department of Chemistry, Tsinghua University, Beijing 100084, China}

\title{LOCAL: A Locality-based Active Learning Framework for Predicting the Stability of  Dual-Atom Catalysts}

\begin{document}

\begin{abstract}
    Dual-atom catalysts supported on nitrogen-doped graphene (DAC/NG) are emerging as a family of promising catalysts that can overcome intrinsic limitations of single-atom catalysts. However, comprehensive assessment of their structural stability is prohibitively demanding due to a vast local configurational space. Here we introduce LOCAL, a locality-based framework that combines graph convolutional networks with active learning to efficiently predict DAC/NG stability by leveraging chemically intuitive locality quantified by crystal orbital Hamilton population analysis. We demonstrate the effectiveness of LOCAL over a comprehensive dataset of 611,648 DAC/NG structures, achieving a test mean absolute error of 0.15~eV while invoking density functional theory calculations for only 16,704 structures (2.7\% of the dataset). Thus, LOCAL enables efficient and accurate construction of phase diagrams for DAC/NG across diverse compositions reciprocally validated with experimentally synthesized configurations for representative systems. Our framework composes an essential methodology for accelerating the discovery and optimization of high-performance complex catalysts.
\end{abstract}

\section{Introduction}

The rational design of high-performance catalysts is essential for advancing sustainable energy, environmental, and chemical engineering. Single-atom catalysts (SACs) present a design framework that offers benefits including maximal atomic utilization and well-defined active sites \cite{Qiao2011, Yang2013, Wang2018, Liu2018, Kaiser2020, Zhuo2020, Li2020}, but they may suffer from inherent limitations—namely, a single metal center and limited coordination environments—which may restrict both their catalytic versatility and activity toward complex reactions \cite{Ma2018, Liu2018NCommun, Jiao2019, Pan2020, Li2022}.

Dual-atom catalysts (DACs), which feature two closely positioned metal atoms anchored on suitable substrates, provide a promising alternative  \cite{wang2017design, li2018synergetic, wang2018synergistic, xiao2018identification, zhang2018coordination}. When supported on nitrogen-doped graphene (NG), DAC/NG systems exhibit superior performance owing to synergistic metal–metal interactions, diverse coordination configurations, and rich electronic structures \cite{yan2017bottom, chu2021neighboring, tian2021dual, zhao2023homonuclear, sun2025dual}. However, the local complexity and compositional diversity of DAC/NG systems make the comprehensive evaluation of their stability an immense challenge, particularly when considering the vast configuration space introduced by different metal combinations, graphene defect patterns, and coordination atoms \cite{Xu2021}. Experimental characterization techniques such as X-ray absorption spectroscopy provide limited resolution for precise atomic-scale structure determination \cite{Ravel:ph5155}, while theoretical modeling based on first-principles methods such as density functional theory (DFT) is prohibitively expensive for exhaustive sampling of all possible DAC configurations.

To overcome this challenge, we propose a locality-based active learning (LOCAL) framework that integrates chemical intuition and active learning to efficiently predict the stability of DAC/NG directly from raw unrelaxed structures, thus bypassing expensive computations. LOCAL comprises two graph convolutional network (GCN) models \cite{Breiman2001, KipfWelling2016, Scarselli2009, Schutt2017, Zhang2019, Glorot2011, Kingma2014, KipfWellingVGAE2016, Hamilton2017, Velickovic2017, Vaswani2017, Han2021, Rampasek2022, Yun2019}: POS2COHP, which estimates local bonding strengths characterized by integrated crystal orbital Hamilton population (ICOHP) values from raw structures, and Graph2E, which predicts the stability energies of active sites using both structural and local bonding information. More importantly, LOCAL employs an active learning scheme that iteratively identifies and selects the worst-described and thus most informative structures for DFT labeling based on chemically intuitive locality characterized by ICOHP, thereby minimizing computational cost while maximizing model performance.

Powered by this framework, LOCAL achieves a test mean absolute error (MAE) of 0.15 eV with invoking DFT calculations for only a small fraction (2.7\%)—16,704 structures—of the dataset, which underwrites its effectiveness—strong performance on these worst-case points increases confidence in predictions across ~600,000 additional raw structures. This enabled an efficient, model-guided survey of the full configuration space and the construction of phase diagrams for all types of DAC/NG considered in our dataset, among which, the predicted stable structures of DAC/NG including Ru--Rh (this work), Co--Ni \cite{Li2023} (previously synthesized by the group of one of authors), Fe--Fe \cite{zhang2020high, zhao2024cascade} (a well validated benchmark), and additional cases \cite{yang2021, cui2022engineering, zhu2022regulating, zhu2023heterogeneous, yue2024tailoring},  agree with experimental characterizations, demonstrating LOCAL’s reliability for discovering stable, experimentally feasible DACs. The dataset and LOCAL framework as well as the predicted phase diagrams were publicly available at www.localenergy.science.

Our results demonstrate LOCAL's efficacy as a powerful tool for predicting the stability of DAC/NG at scale, significantly accelerating the exploration and design of high-performance catalysts. By combining local structural details with active learning, LOCAL is particularly suited to systems with high local variability and modest global diversity, offering strong potential for advancing next-generation catalytic material discovery with broad applicability to various catalyst families beyond DAC/NG.

\section*{Results and Discussion}
\subsection*{The LOCAL Framework}

\begin{figure}[htbp]
\centering
\includegraphics[width=0.99\textwidth]{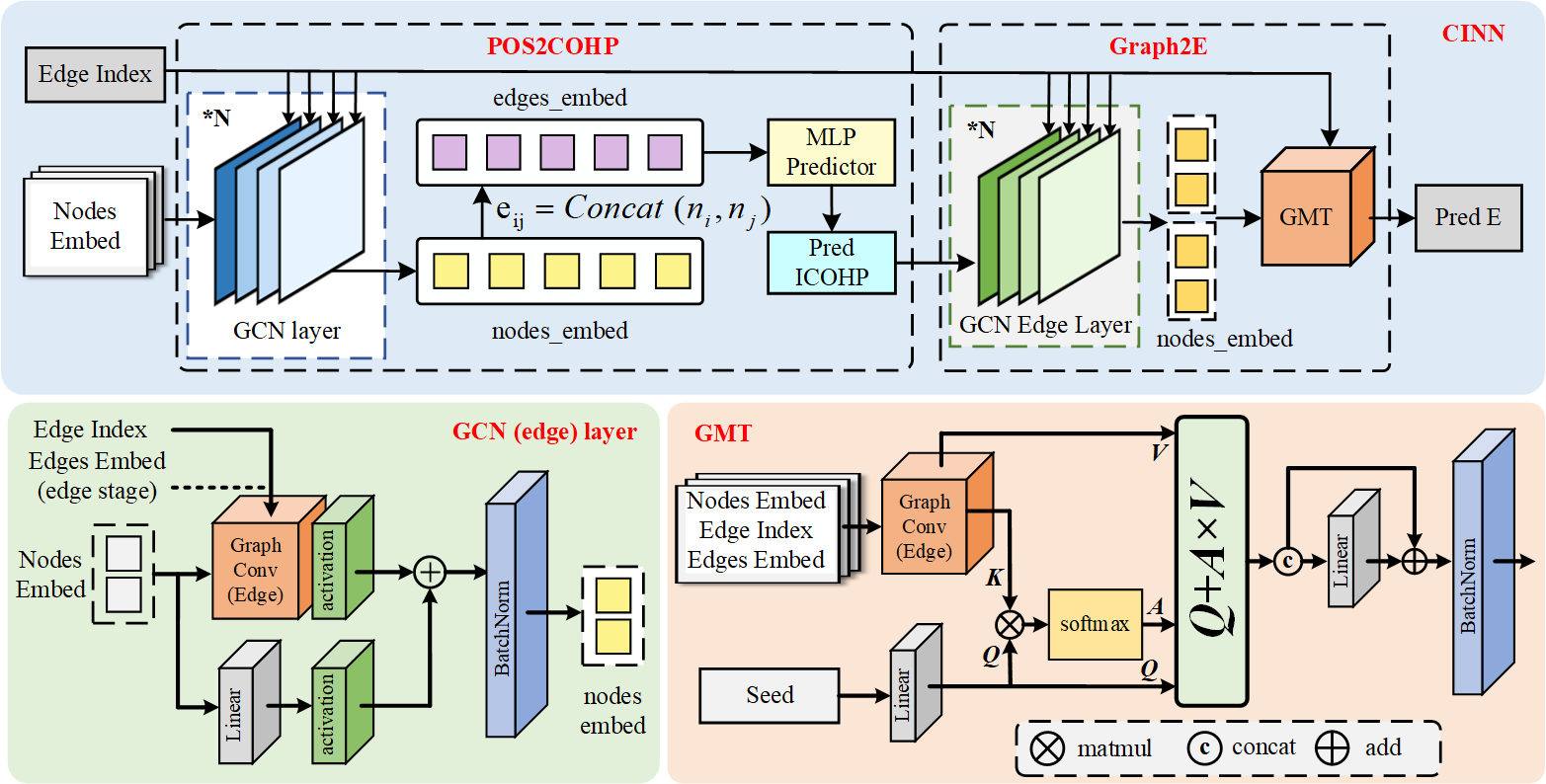}
\caption{The CINN architecture of LOCAL.}
\label{fig:cinn}
\end{figure}

LOCAL leverages a hierarchical chemistry-informed neural network (CINN) architecture as illustrated in Fig.~\ref{fig:cinn}, which integrates two synergistic GCN models, POS2COHP and Graph2E.

The first component, POS2COHP, is designed to predict local bonding strengths by estimating ICOHP \cite{steinberg2018crystal, nelson2020lobster} values directly from the topological structure of unrelaxed raw structures. Each DAC/NG structure is represented as a graph, where atoms are treated as nodes and edges are not explicitly embedded but instead inferred through local coordination patterns. Node features are encoded using one-hot vectors corresponding to element types, capturing the atomic identity and local environment. These node embeddings are processed by graph convolutional layers to learn chemical representations. Metal-ligand node pairs are then extracted and passed through a multi-layer perceptron (MLP) to regress the corresponding ICOHP values, enabling bond strength prediction without optimizing geometries or predefining edge information.

The predicted ICOHP values are subsequently used as edge attributes in the second model, Graph2E, which predicts the target stability energy of a structure (a detailed definition is provided later). Unlike POS2COHP that relies solely on node features, Graph2E explicitly incorporates edge embeddings to represent local bonding strengths, enabling a more chemistry-informed message-passing process. A key component of Graph2E is the transformer-based pooling layer \cite{pablo2023fast,dong2023multi,hu2020heterogeneous,fey2019fast}, which aggregates node-level representations into a fixed-size, high-dimensional global structure embedding. This layer not only condenses the graph information into a compact vector but also learns attention-based relationships between all node pairs, effectively modeling both local interactions and global structural context. The resulting embedding is then passed through a MLP to predict the target stability energy. Additionally, the transformer pooling layer provides an uncertainty estimate, serving as a decoder for active learning and enabling the model to guide data acquisition in an informed and data-efficient manner. (More details on the model architecture and training process can be found in the Methods section.)

\begin{figure}[htbp]
\centering
\includegraphics[width=0.99\textwidth]{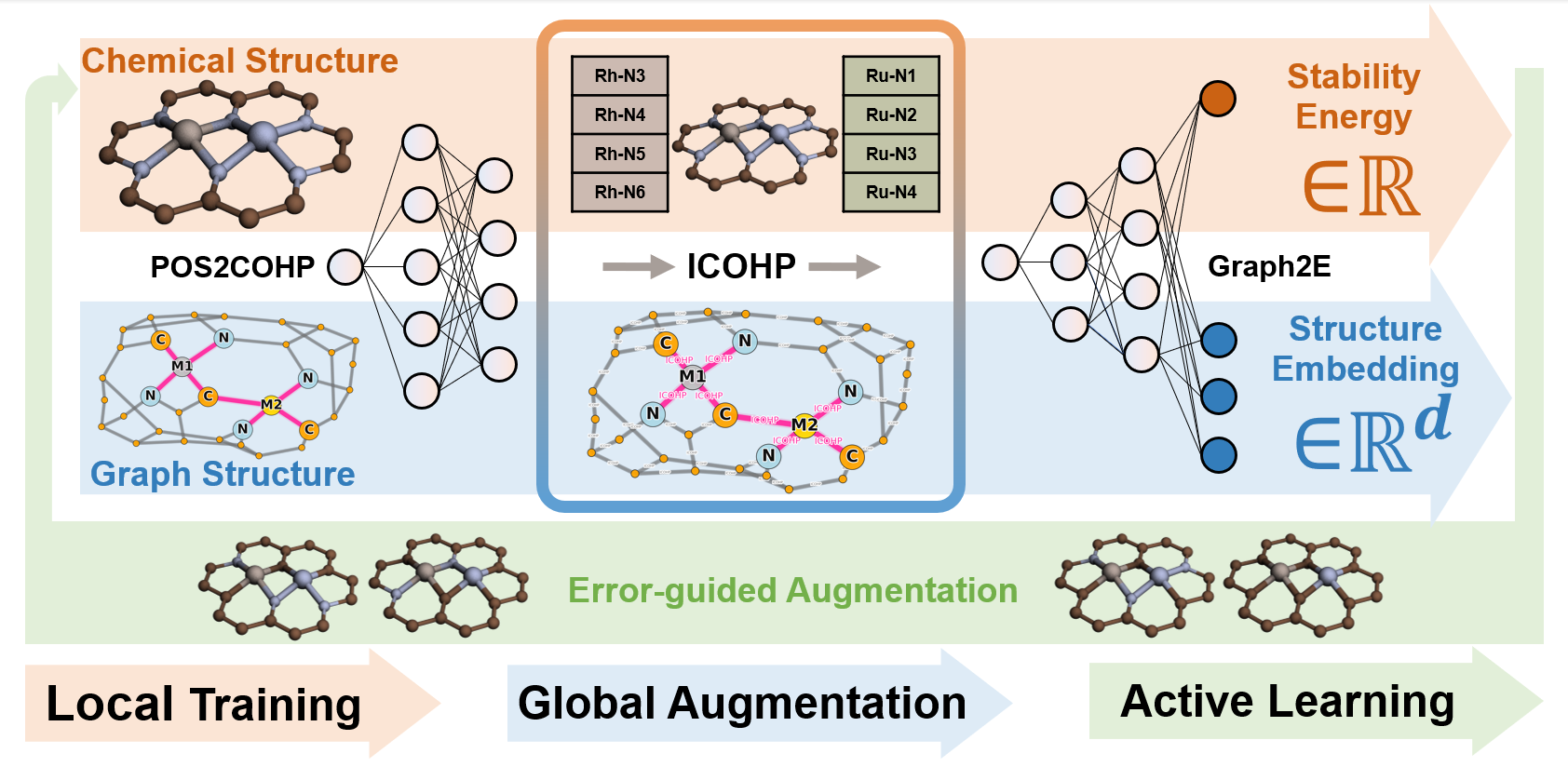}
\caption{The overall workflow of LOCAL.}
\label{fig:workflow}
\end{figure}

Together, POS2COHP and Graph2E form the core of the LOCAL framework, enabling prediction of local bonding strengths and target stability energy directly from an initial structural graph. To effectively apply these models across large-scale datasets, LOCAL adopts a three-stage iterative workflow composed of Local Training, Global Augmentation, and Active Learning, as summarized in Fig.~\ref{fig:workflow}.

In the Local Training stage, a carefully selected subset (details are provided in the next section) of DAC/NG structures (split into training and validation sets by a ratio of 9:1) with DFT-labeled ICOHP values and target total energies is used to train the POS2COHP and Graph2E models, during which POS2COHP learns to predict the required ICOHPs from the topological graph representation of a structure, while Graph2E is trained to subsequently predict the target stability energy using both the original graph and the predicted ICOHPs. After training, the prediction errors on the entire labeled subset are evaluated and ranked, providing a foundation for the error-based sampling in the subsequent Active Learning loops.

In the Global Augmentation stage, the trained POS2COHP and Graph2E models are synergistically applied to the full unlabeled DAC/NG dataset. POS2COHP first predicts the ICOHP values for each structure, which are then used as edge attributes in Graph2E. Rather than predicting energies at this stage, Graph2E is used to extract structure-level embeddings from the penultimate layer of the network. These high-dimensional embeddings, which encode both topological and local bonding information, are organized into a vector space using a KDTree \cite{ram2019revisiting, pedregosa2011scikit} structure to enable efficient similarity and density evaluation across the dataset.

In the Active Learning stage, data points with the highest prediction errors (top 10\%, identified from the Local Training stage) are selected as seed structures. For each seed, a variable number of structurally similar neighbors (up to three) are identified using Euclidean distance in the latent embedding space, defined as

\[
d_{x,y} = \sqrt{\sum_{i=1}^{D}(x_i - y_i)^2}
\]

where \(x\) and \(y\) denote the high-dimensional embeddings of two structures. The newly selected neighbor structures are then evaluated using DFT calculations (details are provided in the Methods section), and their corresponding labels (ICOHP values and target total energies) are added to the training set. This iterative process is repeated until convergence, i.e., when the averaged prediction error of the top 10\% highest-error seed structures falls below 0.20~eV, ensuring that even the model’s worst-case predictions are within the mean absolute deviation by the underlying DFT method per se used for the dataset construction \cite{becke1993density, Heyd2003}.

Consequently, LOCAL integrates chemistry-informed graph representations, scalable similarity evaluation, and uncertainty-guided data acquisition to enable high-fidelity energy prediction across a vast configurational space. Its modular design and data-efficient learning process make it particularly well-suited for accelerating large-scale materials discovery.

\subsection*{Construction of DAC/NG Dataset}

DAC/NG have emerged as a promising family of catalytic materials with diverse coordination environments and synergistic metal–metal interactions, which, however, render a vast number of possible local configurations that pose an immense challenge for understanding and optimizing their catalytic performance. This structural complexity—stemming from variations in metal combination, coordination atoms (C or N), and graphene defect patterns—makes DAC/NG systems both scientifically rich and computationally demanding. These same characteristics also make them ideally suited for the LOCAL framework to evaluate, whose design explicitly targets systems with large local variability but limited global diversity. As such, the DAC/NG dataset was constructed to provide a chemically diverse benchmark for testing LOCAL’s capability in capturing local bonding patterns and predicting stability across an extensive configurational space.

\begin{figure}[htbp]
\centering
\includegraphics[width=0.99\textwidth]{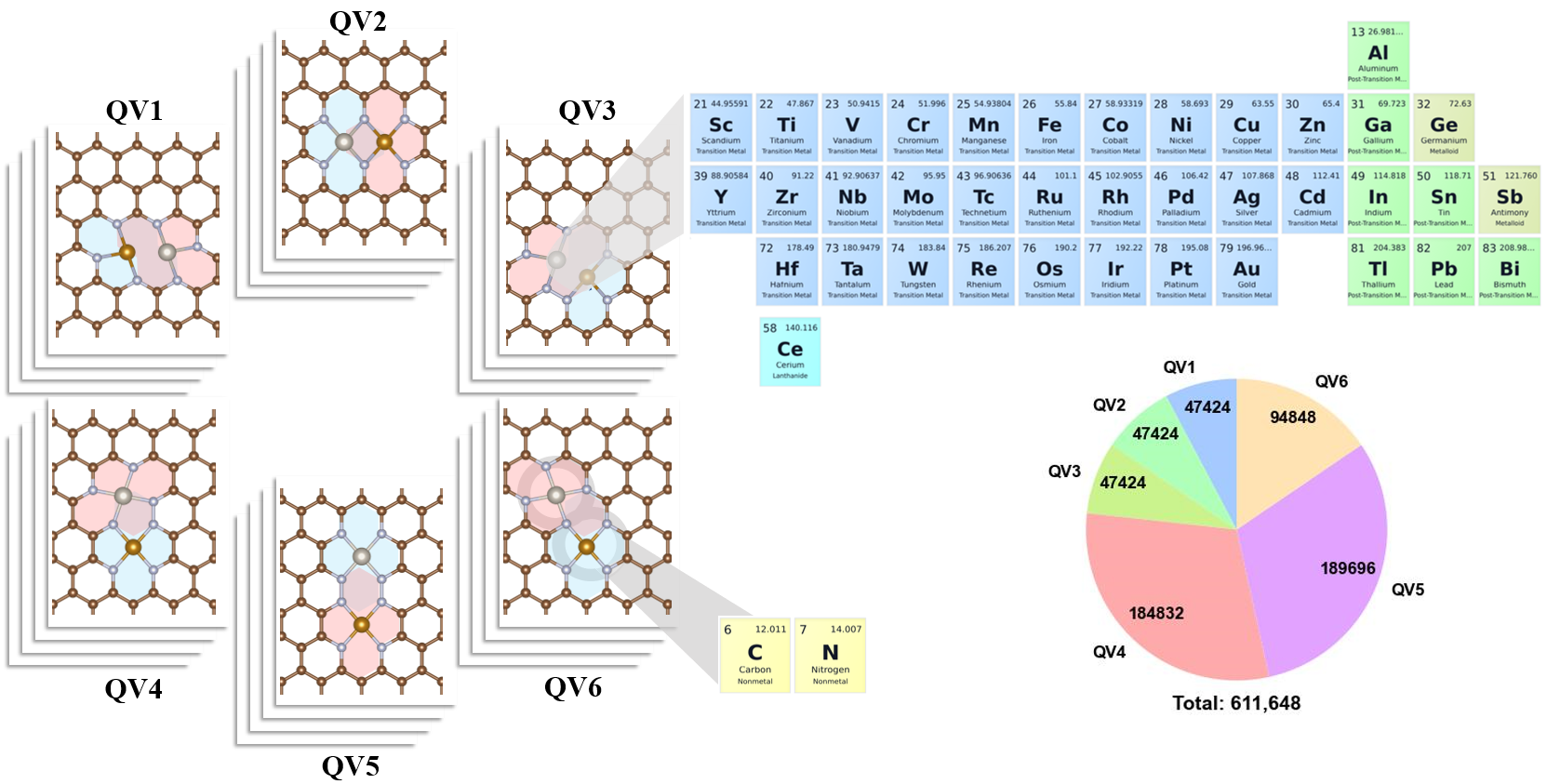}
\caption{Construction of the DAC/NG dataset.}
\label{fig:dataset_construction}
\end{figure}

To construct the DAC/NG dataset, we started with the representative quadra-vacancy (QV) sites in graphene capable of hosting diverse DAC configurations, which are a natural extension of the well-established SAC@NG motifs—formed by removing two adjacent carbon atoms—into dual-atom anchoring environments. To ensure structural feasibility and favorable metal–metal bonding, we merged two SAC-anchoring vacancies with at least one shared hexagon, and this results in six types of QV sites (denoted as QV1–QV6) as shown in Fig.~\ref{fig:dataset_construction}. Next, we populated these QV sites with metal pairs. Considering both transition metals and catalytically relevant main group metal elements, a total of 38 metals were selected. This includes most transition metals from Periods 4–6 (excluding La and Hg), one lanthanide (Ce), and main group p-block elements up to Bi. All possible bimetallic combinations were considered, including both homonuclear and heteronuclear pairs, yielding 741 unique metal pairs. Each pair was then placed into each of the six QV patterns, resulting in 5149 unique metal–defect combinations.

Besides, the DAC/NG systems often incorporate nitrogen doping at defect edges, so we explicitly considered both carbon (C) and nitrogen (N) coordination environments at each metal site. Each QV site permits multiple C/N permutations around the metal, yielding a total of 611,648 unique DAC/NG structures. These subtle variations in coordination, while critical to catalytic performance, are often indistinguishable by conventional experimental characterization techniques. Notably, nitrogen coordination generally increases metal binding strength, as reflected by ICOHP values. Thus, these bond-strength metrics were later used as model descriptors, allowing LOCAL to infer stability without requiring full DFT calculations.

To identify each DAC/NG structure, we employed a systematic, human- and machine-readable naming convention. Each DAC/NG ID, QV\#\_Nindex\_Metals, consists of three components, where QV\# denotes the quadra-vacancy pattern (QV1–QV6), Nindex specifies the nitrogen-doping configuration by indicating which N atoms are replaced with C atoms (with the index defined separately for each QV pattern; if no N atoms are replaced, we use \textit{null} to indicate this), and Metals encodes the two metal elements (with ordering preserved according to the QV pattern). This scheme ensures traceability and consistency across the dataset. (Further details are provided in the SI.)

Conventional adsorption energy calculations for characterizing the stability of dual-atom metal sites become challenging in the presence of multiple types of vacancies due to ambiguous reference energies. To address this, we define the stability energy as
\[
E_\text{target} = E_\text{total}^{\text{DAC/NG}} - n_\text{C} \mu_\text{C} - n_\text{N} \mu_\text{N} - \mu_\text{M1} - \mu_\text{M2}
\]
where \(n_\text{C,N}\) is the number of C or N atoms in DAC/NG and each \(\mu\) term denotes the chemical potential of element defined by its reference as follows. \(\mu_{\mathrm{C}}\) is the chemical potential of C in graphene, \(\mu_{\mathrm{N}}\) is obtained by chemical potential of $\mathrm{N_2}$ in vaccum, and \(\mu_{\mathrm{M1}}, \mu_{\mathrm{M2}}\) are the bulk metal chemical potentials. This metric approximates the relative stability across configurations with universal references.

To accelerate the stability prediction directly from unrelaxed raw structures, we computed ICOHP values between each metal atom and its coordinating C/N atoms, effectively capturing local bonding strengths. These values serve as bonding fingerprints that allow the model to approximate relaxed geometries without expensive DFT relaxation. The number of coordinating neighbors varies by QV pattern: QV2 and QV3 involve three neighbors per metal (six ICOHP values per structure), while QV1, QV4, QV5, and QV6 involve four neighbors per metal (eight ICOHP values per structure).

\begin{table}[htbp]
\centering
\caption{Expansion of the DFT-labeled subset and performance by LOCAL in the active learning loops.}
\label{tab:three_loops}
\begin{tabular}{lccc}
\toprule
\textbf{Loop} & \makecell{\textbf{Number of DAC/NG} \\ \textbf{in the subset}} & \textbf{Testing MAE (eV)} & \makecell{\textbf{MAE (eV) of} \\ \textbf{the top 10\% errors}} \\
\midrule
0 & 13,988 & 0.159 & 0.215 \\
1 & 15,166 & 0.135 & 0.229 \\
2 & 16,704 & 0.153 & 0.184 \\
\bottomrule
\end{tabular}
\end{table}

To label the full DAC/NG dataset with target energies, we relied on the LOCAL framework, which was explicitly designed as an iterative active learning system to identify and select informative structures for local training and global evaluation, as summarized in Table~\ref{tab:three_loops}. In the initial round (Loop0), a subset of 13,988 structures was constructed as the initial labeled subset, which was curated to include chemically diverse and structurally representative configurations, particularly the boundary cases where metal atoms are exclusively coordinated by either carbon or nitrogen atoms, with additional samples randomly drawn across various QV patterns and C/N permutations to enhance generalizability. Then, training and validation splits of this labeled subset were generated to train the POS2COHP and Graph2E models, after which the whole model was used to predict target energies for all structures in the labeled subset, and the prediction errors were evaluated and ranked.

In the subsequent iterations after the initial loop, LOCAL is designed to focus on expanding the labeled dataset by targeting regions of high model uncertainty. Structures exhibiting the largest prediction errors (top 10\%) in the preceding loop are selected as seed points, and a KDTree built on latent embeddings (from the penultimate layer of the Graph2E model) is employed to identify up to three nearest neighbors for each seed structure based on Euclidean distance in the embedding space. These neighbors were then added to the labeled dataset, enriching its chemical diversity and structural representability in a data-efficient manner.

This error-driven and embedding-guided expansion of the labeled subset was repeated in Loops 1 and 2 shown in Table~\ref{tab:three_loops}, adding 1,178 and 1,537 new structures, respectively. The testing MAE decreases from 0.159 eV in Loop 0 to 0.135 eV in Loop 1, before slightly increasing to 0.153 eV in Loop 2. Importantly, the MAE within the top 10\% worst-predicted structures also decreases and remains below the 0.2~eV threshold we set for early stopping, indicating improved model robustness in structurally complex regions.

After this convergence in the prediction performance, the trained LOCAL model was used to label the entire DAC/NG dataset. This enabled large-scale analysis of stability trends across a wide range of metal pairs and provided a foundation for deriving phase diagrams in the resulting chemical space.

\subsection*{Stability Trends and Phase Diagrams of DAC/NG Predicted by LOCAL}

Given that the dataset contains over 600,000 unique structures, analyzing individual cases is neither practical nor informative. Instead, we focus on identifying chemically informative trends from the statistical distribution of stability energies across different QV patterns and metal pair combinations. These aggregate patterns provide insights into structure–stability relationships that can guide the rational design of DAC/NG catalysts.

\begin{figure}[htbp]
    \centering
    \includegraphics[width=0.99\textwidth]{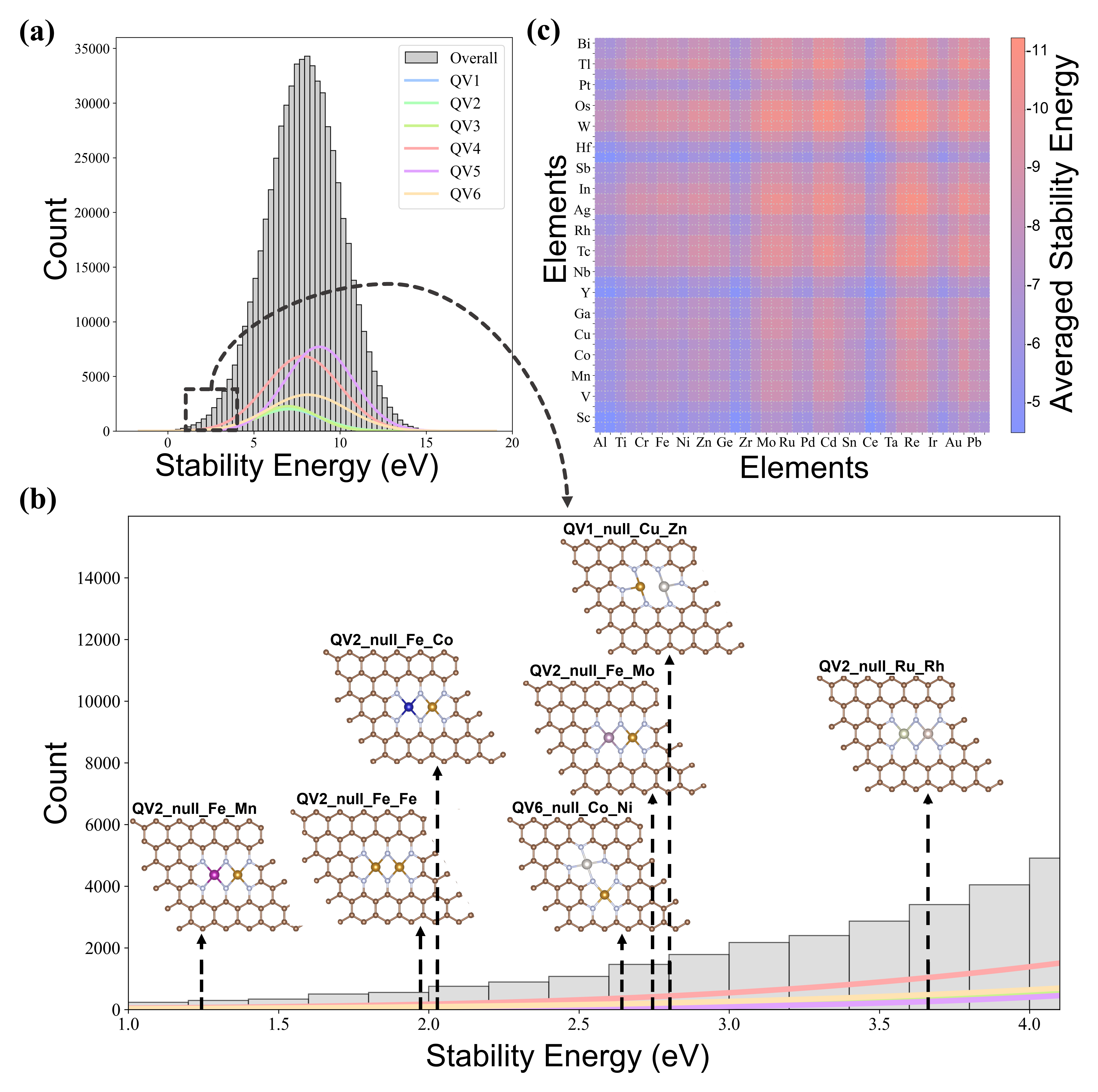}
    \caption{(a) Distribution of predicted stability energies for the DAC/NG dataset.
    (b) Zoomed-in view of the 1–4 eV window in (a), highlighting finer details in the low-energy region with dashed lines marking the experimentally synthesized and effective DAC/NG structures.
    (c) Heatmap of averaged stability energies across all metal pair combinations to identify the element-dependent trends.}
    \label{fig:energydistribution}
\end{figure}

Fig.~\ref{fig:energydistribution}a shows the distribution of predicted target energies, spanning a wide range from -1.65~eV to 19.03~eV and forming a Gaussian-like distribution, which reflects the statistical nature of structural diversity across the dataset. For DAC/NG to serve as high-performing catalysts, overly stable structures may fail to activate reactant molecules, while overly unstable structures are prone to drastic structural evolution under reaction conditions. Therefore, it is essential for rational design of catalysts to identify candidate DAC/NG structures within an intermediate stability energy window, and we used experimentally synthesized and effective DAC/NGs as references to locate the relevant 1–4~eV window, which encompasses the majority of practically accessible structures. In particular, the stability energies of seven experimentally validated DAC/NGs fall within this range, as highlighted in Fig.~\ref{fig:energydistribution}b: QV2\_null\_Mn\_Fe (1.243~eV), QV2\_null\_Fe\_Fe (1.973~eV), QV2\_null\_Fe\_Co (2.028~eV), QV6\_null\_Co\_Ni (2.642~eV), QV2\_null\_Fe\_Mo (2.745~eV), QV1\_null\_Cu\_Zn (2.802~eV), and QV2\_null\_Ru\_Rh (3.662~eV). 

To further analyze the structure–stability relationships, we stratified the dataset by QV patterns. QV1, QV2, and QV3 exhibit relatively moderate and compact energy distributions. Notably, QV2 presents the lowest-energy structure (QV2\_null\_Sc\_Sc), and QV3 shows the narrowest distribution, likely due to enhanced stability by compact geometries. The enhanced stability of these patterns may stem from shorter metal–metal distances in QV1–3, which promote direct metal–metal bonding and contribute to stronger overall binding. In contrast, QV4, QV5, and QV6 feature broader energy spreads and higher variances, reflecting increased coordination flexibility and diversity. QV5, in particular, delivers the highest-energy structure (QV5\_01234567\_Re\_Pb at 19.03~eV), likely owing to its larger void and looser bonding environment.

Besides, Fig.~\ref{fig:energydistribution}c unveils the dependence of stability on metal elements, which demonstrates pronounced periodicity: along either axis, the averaged stability energy naturally partitions into three regions corresponding to Periods~4, 5, and~6 of the periodic table, respectively; the 4--5 boundary lies between Ge and Y, and the 5--6 boundary between Sb and Ce. As the period increases from 4 to 6, the stability energy generally increases, plausibly because heavier metal elements incur greater steric mismatch with the N-doped vacancies. Within a given period, the stability energy exhibits an oscillating trend, with a pronounced minimum near the Ni--Pd--Pt column, which is likely to arise from their tendency to adopt the square planar coordination with the oxidation state of +2. Notably, in Period~4 an additional local minimum occurs at Mn, possibly resulting from the stability of its half-filled $3d^5$ configuration with the oxidation state of +2.

\begin{figure}[htbp]
\centering
\includegraphics[width=0.99\textwidth]{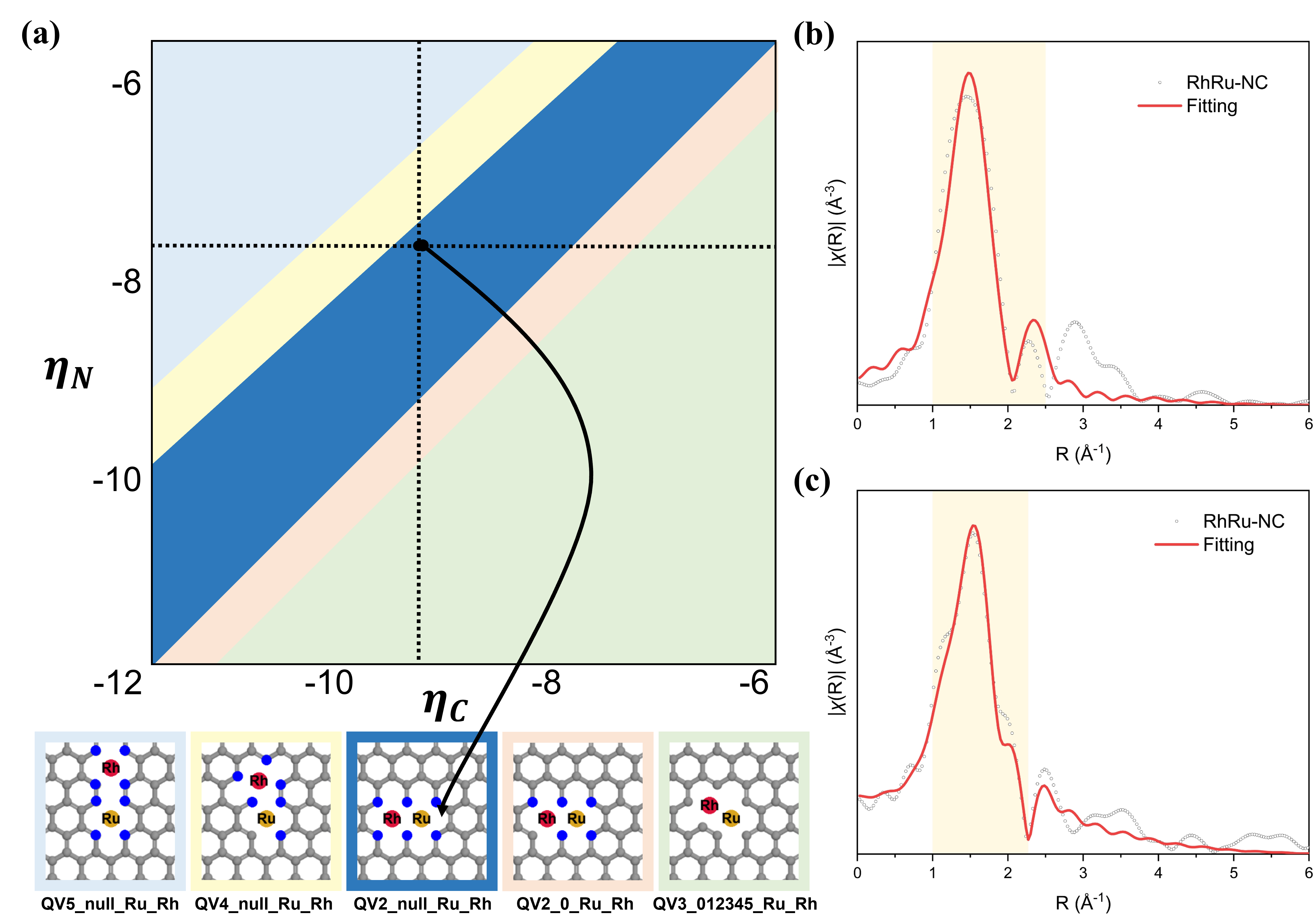}
\caption{Integrated thermodynamic–structural analysis of the Ru–Rh DAC/NG system. (a) Ru–Rh thermodynamic phase diagram with carbon chemical potential \(\eta_{\mathrm{C}}\) as the $x$-axis and nitrogen chemical potential \(\eta_{\mathrm{N}}\) as the $y$-axis. Two dashed guide lines, parallel to the axes, indicate the reference \(\mu_{\mathrm{C}}\) and \(\mu_{\mathrm{N}}\) (graphene for C; g-\(\mathrm{C_3N_4}\) in equilibrium for N; both are at 0 K with no vibrational/\(pV\) corrections; more details are provided in the Methods section); their intersection marks the chosen reference condition at which the dominant phase is identified as QV2\_null\_Ru\_Rh. The five candidate phases are displayed as structural insets at the bottom, arranged from top-left to bottom-right, each outlined in the color corresponding to its representation in the phase diagram. R-space EXAFS fitting of (b) Ru and (c) Rh. In both cases, the coordination number of the first shell is four and that of the Rh–Ru path is one.}
\label{fig:phasediagram}
\end{figure}

These observations align with the experimentally benchmarked stability window of 1–4~eV (Fig.~\ref{fig:energydistribution}b). Among the experimentally realized DAC/NGs in this window, only QV2\_null\_Ru\_Rh contains two fifth-period elements with the highest stability energy of 3.66~eV, whereas the others are composed of fourth-period metals. Overall, both atomic size and electronic structure jointly govern the stability of DAC/NG, offering practical guidelines for selecting metal combinations with enhanced experimental feasibility. Nevertheless, the global analysis above averages out the influence of C and N coordinations; in contrast, the phase diagram for each metal combination recovers this fine factor, as illustrated in Fig.~\ref{fig:phasediagram}a.

Fig.~\ref{fig:phasediagram} integrates LOCAL prediction with experimental validation for Ru–Rh DAC/NG. Fig.~\ref{fig:phasediagram}a shows the LOCAL-predicted phase diagram of Ru–Rh DAC/NG, which indicates the most stable structure under varying chemical potentials of C and N. As the chemical potential of N in the environment decreases, the N content in the most stable configuration decreases, resulting in transition from N-rich to C-rich coordination. We further used this phase diagram to guide our experimental synthesis, under the conditions of which (characterized by chemical potentials of C and N), the phase diagram indicates the most stable structure to be QV2\_null\_Ru\_Rh, consistent with the experimental characterization as discussed below. While many thermodynamically accessible structures are identified for each metal pair, only a limited number of them emerge as dominant under practical chemical conditions. This highlights the value of phase diagrams in narrowing down candidates that are thermodynamically feasible for synthesis. It should be noted, however, that these results assumed thermodynamic equilibrium, while kinetic factors—which may inhibit certain transitions or stabilize metastable states—are not considered in this analysis.

Figs.~\ref{fig:phasediagram}b and \ref{fig:phasediagram}c present the R-space EXAFS fitting for Ru and Rh, respectively. In both cases, the first coordination shell has a coordination number of 4, and there is an additional Ru--Rh path with a coordination number of 1,  supporting a dual-atom configuration. This observation is consistent with the most stable QV2\_null\_Ru\_Rh DAC/NG structure identified with the LOCAL-predicted phase diagram. However, because the first-shell EXAFS cannot unambiguously distinguish N from C neighbors within the experimental uncertainty, the experiment establishes the coordination numbers (4 for Ru/Rh and 1 for Ru--Rh) but not the elemental identity of the first-shell ligands. To further validate this, we performed DFT calculations on all four-coordinated QV2 variants with different N/C contents (see the Supplementary Notes for additional details), and the resulting lowest-energy model also agrees with the phase-diagram prediction by LOCAL.

Beyond this case study, earlier experimental synthesis of Co--Ni \cite{Li2023}, Fe--Mn \cite{cui2022engineering}, and Fe--Mo \cite{zhu2022regulating} by the group of one of authors also match the phase-diagram predictions by LOCAL, further demonstrating the robustness of our approach. While our analysis focuses on thermodynamic equilibrium (and thus does not explicitly account for kinetic barriers or metastability), the resulting phase diagrams are effective for narrowing down synthesis-feasible candidates. More broadly, we generated thermodynamic phase diagrams for all 741 unique bimetallic combinations, enabling systematic comparisons across coordination motifs. To facilitate exploration and application of these phase diagrams, we provide an interactive platform at \url{www.localenergy.science/phasediagrams.html}, where readers can select metal pairs, vary $\mu_{\mathrm{C}}$ and $\mu_{\mathrm{N}}$, and visualize the predicted dominant structures. Taken together, this integrative workflow—combining graph-based energy prediction with thermodynamic phase-stability analysis—offers a practical route to identify DAC/NG configurations that are thermodynamically viable under experimental conditions and facilitate their rational design.

\section*{Conclusions}

In summary, we introduced LOCAL, a locality-based active-learning framework for predicting the atomic-scale stability of materials characterized by large variability of local structure but limited diversity of global structure. By integrating two GCN models, POS2COHP for local bond-strength estimation and Graph2E for global energy prediction, LOCAL leverages chemically informed graph representations and uncertainty-guided data acquisition to achieve high predictive accuracy (MAE = 0.15~eV) directly from unrelaxed raw atomic configurations.

We demonstrated the effectiveness of LOCAL by investigating the stability of DAC/NG that has large local variability but limited global diversity. In doing so, we constructed a comprehensive dataset of 611,648 DAC/NG structures covering 38 metals, multiple defect motifs, and diverse coordination environments. The application of LOCAL to this dataset delivered efficient DFT labeling, large-scale stability prediction, and systematic construction of thermodynamic phase diagrams for 741 bimetallic combinations, which were reciprocally validated with experimental synthesis and characterization in representative cases, and thus can provide extensive guidelines for rational design and experimental synthesis of DAC/NG.

LOCAL is inherently general and can be applied to a wide range of materials beyond DAC/NG and to the studies of activity by investigating reaction intermediates that are instrinsically local. By leveraging local chemical and structural information, LOCAL enables rapid, scalable evaluation of stability and activity across vast configurational spaces, bridging the gap between computational prediction and experimental realization. Its modular design and transferability make it a broadly applicable framework for accelerating the discovery and optimization of next-generation, high-performance catalytic materials.

\section{Methods}

\subsection*{DFT Calculations}

All DFT calculations were performed with the projector-augmented wave (PAW) method as implemented in \textsc{VASP} \cite{kresse1993ab, kresse1994norm, kresse1996software, kresse1996efficient, kresse1999ultrasoft}. The approximation of Perdew–Burke–Ernzerhof (PBE) \cite{ernzerhof1999assessment} for the exchange–correlation functional was employed, and a plane-wave kinetic-energy cutoff of 500~eV was used throughout.

Each DAC/NG model was built by embedding the dual-metal motif in a graphene supercell with at least 15~\AA\ of vacuum normal to the sheet to suppress spurious interlayer interactions; dipole corrections were applied along the vacuum direction. The in-plane Brillouin zone was sampled with a $\Gamma$-centered $3\times3\times1$ Monkhorst–Pack mesh for all supercells; this mesh was validated to converge total energies to within 1~meV per atom for a representative subset. Spin polarization was enabled for all structures; initial magnetic moments on the transition-metal sites were set according to high-spin Hund-rule guesses and allowed to relax self-consistently; the electronic structures were converged with the total-energy change to be $<\!10^{-5}$~eV. We fixed the lattice vectors to those of pristine graphene and relaxed all internal coordinates until forces were $<\!0.03$~eV\,\AA$^{-1}$.

For a given structure $i$, the resulting total energy $E_{\rm total}^i$ was combined with elemental chemical potentials to calculate the stability energy used by LOCAL,
\[
E_\text{target}^i = E_\text{total}^{i} - n_\text{C}^i\,\mu_\text{C} - n_\text{N}^i\,\mu_\text{N} - \mu_\text{M1}^i - \mu_\text{M2}^i,
\]
where $n_\text{C}^i$ and $n_\text{N}^i$ are the numbers of C and N atoms. The carbon reference $\mu_\text{C}$ was taken from pristine graphene. The nitrogen reference $\mu_\text{N}$ was taken as one half of the total energy of an isolated $\mathrm{N_2}$ molecule placed in a large periodic box (vacuum $>\!15$–20~\AA). Because $\mu_\text{C}$ and $\mu_\text{N}$ enter as thermodynamic variables in Eqs.~(4)–(6), any constant shift of $\mu_\text{C}$ or $\mu_\text{N}$ (e.g., using a g-$\mathrm{C_3N_4}$–derived $\mu_\text{N}$ instead of $\tfrac{1}{2}E_\mathrm{tot}(\mathrm{N_2})$) rigidly translates the ($\mu_\text{C}$,$\mu_\text{N}$) axes but does not change phase boundaries or the identity of the stable phase at a given ($\mu_\text{C}$,$\mu_\text{N}$). Metal chemical potentials $\mu_{\rm M1}$ and $\mu_{\rm M2}$ were obtained from the lowest-energy elemental bulk phases using the same cutoff and comparable $k$-point densities to the slab models; energies were normalized per atom.

\subsection*{ICOHP Calculations and Bond Assignment}

Local bond strengths were quantified by the ICOHP analysis using \textsc{LOBSTER} \cite{Nelson2020, Maintz2016} on top of \textsc{VASP} wavefunctions. In line with our production workflow, ICOHPs were evaluated from single-point electronic calculations on the relaxed geometries. Spin polarization was retained; spin-resolved COHP curves were summed to obtain total COHPs.

We used LOBSTER’s recommended minimal basis sets: C($2s\,2p$), N($2s\,2p$); 3$d$ metals ($3d\,4s\,4p$), 4$d$ metals ($4d\,5s\,5p$), and 5$d$ metals ($5d\,6s\,6p$). COHP curves were computed on an energy grid spanning at least $[-10,\,10]$~eV relative to $E_F$ with $\geq\!2000$ points. LOBSTER provides orbital-resolved ICOHP contributions between atoms $A$ and $B$ for each orbital pair $(\ell,\ell')$. We summed these channel contributions to obtain the atom–atom ICOHP,
\[
\mathrm{ICOHP}_{A\!-\!B} \;=\; \sum_{\ell,\ell'} \mathrm{ICOHP}_{A(\ell)\!-\!B(\ell')}
\;=\; \int_{-\infty}^{E_F} \mathrm{COHP}_{A\!-\!B}(\varepsilon)\,\mathrm{d}\varepsilon,
\]
using LOBSTER’s sign convention (more negative values indicate stronger bonding). 

We restricted the COHP analysis to first-shell metal–ligand pairs only, i.e., bonds between each metal and its coordinating C or N neighbors dictated by the QV motif. Direct metal–metal COHPs were not included in the feature set. Consistent with the vacancy topology, QV2 and QV3 provide three M–(C/N) neighbors per metal (six per structure), while QV1 and QV4–QV6 provide four per metal (eight per structure). 

\subsection*{Phase‐Diagram Construction}

After obtaining the full set of LOCAL–predicted target energies \(E_{\rm target}^i\) for each DAC/NG structure \(i\), we recall from its definition that 
\begin{equation}
    E_{\rm target}^i=E_{\rm total}^{i}- n_C^i\,\mu_C- n_N^i\,\mu_N- \mu_{M1}^i- \mu_{M2}^{i}
\end{equation} 
where \(E_{\rm total}^{i}\) is the total energy for structure \(i\) and each \(mu\) is taken from its reference state (graphene for C, molecular \ce{N2} gas for N, and the bulk metal phase for \(M_1\) and \(M2\).
Rearranging Eq.(1) gives the total energy \(E_{\rm total}^{i}\) as,
\begin{equation}
    E_{\rm total}^{i} = E_{\rm target}^i + n_C^i\,\mu_C + n_N^i\,\mu_N + \mu_{M1}^i + \mu_{M2}^{i}
\end{equation} 

Taking the carbon and nitrogen chemical potentials as independent variables \(\eta_C\) and \(\eta_N\) in place of \(\mu_C\) and \(\mu_N\), Eq. (1) can be expressed as a plane in the \((\eta_C,\eta_N,E_{\rm target})\) space,
\begin{equation}
    n_C^i\,\eta_C \;+\; n_N^i\,\eta_N \;+\; 1\cdot E_{\rm target}^i \;+\; \underbrace{\bigl(\mu_{M1}^i+\mu_{M2}^i - E_{\rm total}^i\bigr)}_{d_i} \;=\;0
\end{equation} 
Comparing to the standard plane equation \(a x + b y + c z + d = 0\), we identify
\[ x = \eta_C,\quad y = \eta_N,\quad z = E_{\rm target}^i,\quad a = n_C^i,\quad b = n_N^i,\quad c = 1,\quad d = d_i . \]

Building on these definitions, for a fixed metal pair \((M_1,M_2)\) let \(\{i\}\) be the set of candidate structures.  From Eq.\,(3), each \(i\) defines a plane in the \((\eta_C,\eta_N,z)\) space,
\begin{equation}
    z = E_{\rm target}^i(\eta_C,\eta_N)  = -\,n_C^i\,\eta_C -\,n_N^i\,\eta_N -\,d_i
\end{equation} 
At any fixed \((\eta_C,\eta_N)\), the thermodynamically stable phase is the structure \(i\) with the lowest stability energy,
\begin{equation}
    E_{\rm stable}(\eta_C,\eta_N) = \min_{i}\bigl\{E_{\rm target}^i(\eta_C,\eta_N)\bigr\}.
\end{equation} 
Phase boundaries then occur along intersections of two planes,
\begin{equation}
    E_{\rm target}^i(\eta_C,\eta_N) = E_{\rm target}^j(\eta_C,\eta_N) \;\Longrightarrow\; (n_C^i - n_C^j)\,\eta_C + (n_N^i - n_N^j)\,\eta_N + (d_i - d_j) = 0
\end{equation}
which partition the \((\eta_C,\eta_N)\) plane into regions where each \(i\) minimizes \(E_{\rm target}\), yielding the final phase diagram.

\subsection*{Elemental Chemical Potentials from Graphene (C) and g-\(\mathrm{C_3N_4}\) (N)}

To identify the most stable configuration from the phase diagram for the experimental synthesis conditions, we need to determine the corresponding values of carbon and nitrogen chemical potentials. Because this work focuses on N-doped graphene, we take graphene as the carbon reference and infer the nitrogen chemical potential by imposing equilibrium with g-\(\mathrm{C_3N_4}\), the only well-established 2D carbon nitride whose local C/N coordination closely mirrors N-doped graphene, thereby providing a chemically consistent nitrogen reference. Since both references are 2D solids, we neglect vibrational and \(pV\) contributions and take \(G \approx E^{\rm DFT}\). We use 0 K DFT energies with units
\begin{equation}
E^{\rm DFT}_{\mathrm{graphene}}= -9.223\ \mathrm{eV/atom}, 
\qquad
E^{\rm DFT}_{\mathrm{C_3N_4}}= -58.571\ \mathrm{eV/f.u.}
\end{equation}

For getting the carbon chemical potential (graphene reference), we set the carbon chemical potential directly from graphene,
\begin{equation}
\mu_C^{\mathrm{graphene}} = E^{\rm DFT}_{\mathrm{graphene}}
\end{equation}

For calculating the nitrogen chemical potential from g-\(\mathrm{C_3N_4}\), we utilize the equilibrium per formula unit of g-\(\mathrm{C_3N_4}\) that imposes
\[
\mathrm{g\text{-}C_3N_4} \;\rightleftharpoons\; 3\,\mathrm{C}(\text{graphene reference}) + 4\,\mathrm{N}(\text{nitrogen reference}), 
\]
\begin{equation}
E^{\rm DFT}_{\mathrm{C_3N_4}} = 3\,\mu_C + 4\,\mu_N
\end{equation}
which yields
\begin{equation}
\mu_N^{\mathrm{g\text{-}C_3N_4\text{-}ref}} = \frac{1}{4}\!\left(E^{\rm DFT}_{\mathrm{C_3N_4}} - 3\,\mu_C^{\mathrm{graphene}}\right)=-7.726\ \mathrm{eV}
\end{equation}

The final resulting reference values for identifying the most stable configuration under the experimental synthesis conditions in the phase diagram are
\begin{equation}
\eta_C^\circ=\mu_C^{\mathrm{graphene}}=-9.223\ \mathrm{eV}, \qquad \eta_N^\circ=\mu_N^{\mathrm{g\text{-}C_3N_4\text{-}ref}}=-7.726\ \mathrm{eV}.
\end{equation}
So we plot this \((\eta_C^\circ,\eta_N^\circ)\) point in the phase diagram and identify the stable DAC/NG configuration.

\begingroup
\newcommand{\BN}{\mathrm{BN}}
\newcommand{\Dropout}{\mathrm{Dropout}}
\newcommand{\ReLU}{\mathrm{ReLU}}
\newcommand{\Concat}{\mathrm{Concat}}
\newcommand{\GCN}{\mathrm{GCN}}
\newcommand{\GCNedge}{\mathrm{GCN}_{\mathrm{edge}}}

\subsection{Model expressions of the LOCAL framework}

This section gives the mathematical formulations of \texttt{POS2COHP} and \texttt{Graph2E} used in the LOCAL framework.

\paragraph{POS2COHP.}
Let node $i$ have features $h_i^{(k-1)}$ at layer $k\!-\!1$, and denote its neighborhood by $\mathcal{N}(i)$.
For layer $k$, the aggregation and residual terms are
\begin{equation}
m_i^{(k)}=\sum_{j\in \mathcal{N}(i)}\Bigl(W_v^{(k)}\,h_j^{(k-1)}+b_v^{(k)}\Bigr).
\end{equation}
\begin{equation}
r_i^{(k)}=W_r^{(k)}\,h_i^{(k-1)}+b_r^{(k)}.
\end{equation}
After applying ReLU to both terms, we add them and then pass through Dropout and BN to obtain the layer output:
\begin{equation}
h_i^{(k)}=\BN\!\Bigl(\Dropout\!\bigl(\ReLU(m_i^{(k)})+\ReLU(r_i^{(k)})\bigr)\Bigr).
\end{equation}
The first layer uses $h_i^{(0)}=x_i$ (raw node features). A single GCN layer is abstracted as
\begin{equation}
x^{\mathrm{out}}=\GCN_k\!\left(X^{\mathrm{in}},\,\mathcal{G}\right).
\end{equation}
Finally, concatenate the two endpoints’ node outputs from layer $K$ to obtain an edge representation:
\begin{equation}
e_{ij}=\Concat\!\bigl(h_i^{(K)},\,h_j^{(K)}\bigr).
\end{equation}
Map it to the output space (ICOHP) with an MLP:
\begin{equation}
P_t=W_{\mathrm{MLP}}\!\left(\BN\!\left(\ReLU\!\bigl(W_{\mathrm{MLP}}^{\mathrm{hidden}}\,
\Dropout(e_{ij})+b_{\mathrm{MLP}}^{\mathrm{hidden}}\bigr)\right)\right)+b_{\mathrm{MLP}}.
\end{equation}

\paragraph{Graph2E.}
Inputs are node features $X$ and edge attributes $E$. For each node $i$ and its neighbor $j\!\in\!\mathcal{N}(i)$, layer $k$ first applies linear transforms:
\begin{equation}
h_i^{(k-1)'}=W_h^{(k)}\,h_i^{(k-1)}+b_h^{(k)},
\end{equation}
\begin{equation}
e_{ij}^{(k-1)'}=W_e^{(k)}\,e_{ij}^{(k-1)}+b_e^{(k)}.
\end{equation}
The aggregation and residual terms are then
\begin{equation}
m_i^{(k)}=\sum_{j\in\mathcal{N}(i)}\Bigl(h_j^{(k-1)'}+e_{ij}^{(k-1)'}\Bigr),
\end{equation}
\begin{equation}
r_i^{(k)}=W_r^{(k)}\,h_i^{(k-1)'}+b_r^{(k)}.
\end{equation}
After ReLU, Dropout, and BN, the layer output is
\begin{equation}
h_i^{(k)}=\BN\!\Bigl(\Dropout\!\bigl(\ReLU(m_i^{(k)})+\ReLU(r_i^{(k)})\bigr)\Bigr).
\end{equation}
For the first layer, $h_i^{(0)}=x_i$ and $e_{ij}^{(0)}=E_{ij}$ (raw node and edge features). We denote an \texttt{GCN\_edge} layer as
\begin{equation}
x^{\mathrm{out}}=\GCNedge\!\left(X^{\mathrm{in}},\,E,\,\mathcal{G}\right).
\end{equation}

Finally, feed node outputs $X^K$ to a pooling module. Introduce learnable seeds $S$ (queries in PMA).
Let $S$ produce the Query, while Key and Value come from edge-aware graph convolution:
\begin{equation}
Q=W^{Q}S,\qquad
K=\GCNedge^{\text{key}}\!\left(X^{K},E,\mathcal{G}\right),\qquad
V=\GCNedge^{\text{value}}\!\left(X^{K},E,\mathcal{G}\right).
\end{equation}
Split $Q,K,V$ along the last dimension into $H$ heads, each with size
\begin{equation}
d=\frac{d_v}{H}.
\end{equation}
Let the $h$-th head be $Q_h,K_h,V_h$ ($h=1,\dots,H$). Scaled dot-product attention and head output are
\begin{equation}
A_h=\mathrm{softmax}\!\left(\frac{Q_hK_h^{\top}}{d_v}\right),\qquad
O_h=Q_h+A_hV_h.
\end{equation}
Concatenate all heads:
\begin{equation}
O=\Concat(O_1,O_2,\dots,O_H),
\end{equation}
and apply a linear transform, activation, and residual connection to obtain the final output:
\begin{equation}
Z=O+\ReLU\!\bigl(W^{O}(O)\bigr).
\end{equation}

\endgroup

\begin{acknowledgement}
This work was supported by the National Natural Science Foundation of China (22525302), the NSFC Center for Single-Atom Catalysis (22388102), and the National Key Research and Development Project (2022YFA1503000). We are also grateful to the Center of High-Performance Computing at Tsinghua University for providing computational resources.
\end{acknowledgement}

\bibliography{references}

\end{document}